# Rejoinder: Harold Jeffreys's Theory of Probability Revisited

Christian P. Robert, Nicolas Chopin and Judith Rousseau

*Abstract.* We are grateful to all discussants of our re-visitation for their strong support in our enterprise and for their overall agreement with our perspective. Further discussions with them and other leading statisticians showed that the legacy of *Theory of Probability* is alive and lasting.

## 1. ON BERNARDO'S COMMENTS

We cannot but agree with most issues raised by Professor Bernardo, first and foremost the important distinction between testing and estimation. The multidimensional Jeffreys prior (for estimation) is certainly not formally defined within *Theory of Probability* and the multiplication of cases in the book does not help. We alas have no clear explanation as to why most Jeffreys priors produce proper posteriors for all datasets. While Lindley–Jeffreys's paradox may be upsetting (although it mostly highlights the discrepancy between the frequentist and the Bayesian answers) we, however, consider the attempt to create a testing Jeffreys prior in Section 5.2 as an interesting if incomplete attempt, concretized much later by Bayarri and Garcia-Donato (2007). We understand Professor Bernardo's point of view on Bayes factors, but still resist the temptation to throw away this useful tool, as discussed below in conjunction with Professor Lindley's comments. We are nonetheless sympathetic to the intrinsic discrepancy measure as an invariant loss function, even though the necessity of scaling this measure and of selecting the subsequent bound between acceptance and rejection remain strong impediments against adopting this alternative. (The fact that it depends on the sample size $n$ is certainly a major drawback, although one could reasonably object that the bound between acceptance and rejection associated with the Bayes factor should also depend on the sample size.)

## 2. ON GELMAN'S COMMENTS

We first apologize to the authors of Gelman et al. (2001) for not ranking them into the "classics" of our first footnote. This choice was, however, deliberate: we wanted to stop short of comparing the most recent textbooks of the late 1990s (excluding as well Robert, 1994). At a more foundational level, the debate about the choice of a noninformative or of a weakly informative prior is endless, hopeless and possibly fruitless, in that (a) there is no way *a single perfect* noninformative prior can be adopted by one and all except through a formal decision from the community to always use Jeffreys prior as a default (in the same way the Black-and-Scholes formula is used by financial analysts as a common ground, not as a representation of real series); and (b) noninformative and informative priors are not two well-separated categories, they form a continuum. It seems thus more fruitful to try to build measures that assess of the impact of a given prior (or of the variation of a parameter in a family of priors). The debate about complexity is more in line with our views: (a) similar to the notion of a universal noninformative prior, a practical implementation of

*Christian P. Robert is Professor, CEREMADE, Université Paris Dauphine, 75775 Paris cedex 16, France e-mail: xian@ceremade.dauphine.fr. Nicolas Chopin is Professor, CREST-ENSAE, INSEE, 92245 Malakoff cedex, France e-mail: nicolas.chopin@ensae.fr. Judith Rousseau is Professor, CEREMADE, Université Paris Dauphine, 75775 Paris cedex 16, France e-mail: rousseau@ensae.fr.*

*the* Ockham's razor does not exist; and (b) complexity is quite a subjective factor. This is not to say that we reject the Jeffreysian tenet that Bayes factors naturally downweight complex models with limited support from the data, since we support this point, but rather that the support for more complex models should come from the prior or from a loss function, rather than from the complexity-hungry likelihood. (The social scientist attitude that worries about missing some factor could also be questioned as being too optimistic in its belief in models.) Finally, we concede that Bayesian data analysis may force us to move away from the "ideal" standards set by Harold Jeffreys's *Theory of Probability*, including the reliance on the Bayes factor. Bayesian model criticism, indeed a major direction in Gelman et al. (2001), is still in its infancy and would correctly require more emphasis in our papers and in our practice! As put by Professor Gelman, we need to learn more from "the failures of a statistical model's attempt to capture reality."

## 3. ON KASS' COMMENTS

We are grateful to Professor Kass for his comments that follow a talk given during the Harold Jeffreys's *Theory of Probability* anniversary session at the O-Bayes 2009 meeting. There is actually very little we can disagree with in these comments which show a deep and scholarly knowledge of *Theory of Probability* and expose our need to pursue our study of this profound book.

The connection with geometry was bound to be part of Professor Kass' comments and we do agree with the essential feature of looking for orthogonal parameterisation, a point which, in our awkward phrasing, we would relate to the search for a constant information parameterization. We also appreciate the emphasis on Laplace's approximations that permeate the book and provide an early link with Bayesian asymptotics. The epistemological implications of *Theory of Probability* are certainly worth stressing (a point also made by Professor Zellner) if only because Harold Jeffreys was first and foremost a physicist who developed his own statistical tools to deal with his own physics problems. The specific points made by Professor Kass about the nature of statistical models would be worth emphasising during any course in applied and even methodological statistics (as are the central discussions by Erich Lehmann and David Cox in the 1990 volume of this journal). That Bayesian testing, or any kind of testing, remains a source for discussion and further research is clearly illustrated by the number of comments and the variety of proposals on this point. Finally, the lack of decision theory is an issue that we also deplore, in agreement with Professors Bernardo and Lindley as well, if not Professor Zellner.

## 4. ON LINDLEY'S COMMENTS

Besides so kindly contributing to the discussion therein, Professor Lindley patiently and helpfully enlightened us on the construction and contents of *Theory of Probability* during the preparation of the paper. We are therefore deeply indebted to him for sharing so much with us. His comments bring a unique perspective to the discussion, both from historical and foundational viewpoints. As a witness of the early developments of *Theory of Probability*, Professor Lindley exposes the philosophical cum practical reasons for the composition of this book. The point about Section 3.10 and the integration over the sample space was missed in our analysis but is indeed crucial in its link with the likelihood principle that does not appear per se in *Theory of Probability*. Nowadays, this is certainly the most standard example that illustrates how the principle for constructing Jeffreys's priors may violate the likelihood principle (Berger and Wolpert, 1988). (The opposition with deFinetti's perspective is also worth noticing, since they approached Bayesian statistics from fundamentally different perspectives, even though their respective books share the same title.)

The fact that uncertainty must be analyzed in probabilistic terms is certainly a driving force in *Theory of Probability* and a convincing reason to follow Bayesian ways. We completely agree that this formalization is one of Harold Jeffreys's great inputs. Once again, the other fundamental input stressed both by Professor Lindley and ourselves is the complete formalization of a coherent approach to testing via Bayes factors. Professor Lindley is 100% correct in his assessment of the opposition of this view with Popper's and of its persistence (Templeton, 2008): rejecting a model based on its "falsity" is only feasible when considering the available alternatives. That *Theory of Probability* does not directly dwell on decisions is clearly a feature of the time, even though Keynes had opened the way a few years earlier, but this did not prevent a formalization of Bayesian testing procedures that proved itself compatible with "0–1" loss functions, thus showing the insight in Harold Jeffreys's intuitions.

## 5. ON SENN'S COMMENTS

Given the tone of some earlier comments of Professor Senn on the Bayesian paradigm, we must acknowledge our pleasant surprise at his conciliatory tone in these comments. Thankfully, the barbed parody of Harold Jeffreys's most quoted sentence somehow re-establishes the balance! We are quite grateful to Professor Senn for his laudatory remarks, even though we must acknowledge that our copies of Harold Jeffreys's *Theory of Probability* are also full of pencil annotations and question marks, and also that it took two series of lectures to achieve this incomplete state of awareness. We furthermore enjoyed the mention that Harold Jeffreys considered Bayesian significance tests as the most important part of *Theory of Probability*, since this agrees with both Professor Lindley's and our perceptions.

We dearly appreciate the further historical details provided by Professor Senn's comments, particularly in that the exchange between Ronald Fisher and Harold Jeffreys is represented in much less a controversial tone that we could have believed! [The first author also commented on Berger, Bernardo and Sun (2009) about the particular matter of the Law of Succession and so we do not need to repeat the comments here.] Similarly, the confusion about Bernoulli shows how amateurish is our attempt at Science History. We are equally grateful to Professor Senn for pointing out Bartlett's connection, as we must confess we were not even aware of it! When reading Bartlett's comments, we came to realize his contribution to the exclusion of improper priors for Bayes factors, as analyzed in deeper details by Bickel and Ghosh (1990).

## 6. ON ZELLNER'S COMMENTS

Unsurprisingly, Professor Zellner's comments—that he delivered quite enthusiastically during his lecture at O-Bayes 2009—are opening new vistas on *Theory of Probability*, while differing from our analysis on several points. The first issue is that *Theory of Probability* was aimed at scientists at large, while we read it as statisticians. This is unavoidable, given our background and, further, we doubt many non-statisticians would have the time and the will to go through *Theory of Probability*. Unfortunately, most of them seem to eschew modern Bayesian introductions to the benefit of shorter reviews published in their own discipline. We completely agree with Professor Zellner that we failed to understand the historical undercurrents explaining the connection of *Theory of Probability* with the philosophy of science at the time it was written. This is not to say we missed the global impact of *Theory of Probability* on scientific modeling and its definition of induction, a point already stressed by Professor Kass, because it obviously represents the major impact of the book, but the style of the discussions about the axiomatic nature of probability and our lack of background in this area led us to bypass them to focus on the link with modern Bayesian statistics. (Neither does the "proof" of Bayes' theorem strike us as ultimately necessary, once the axiomatic definition of probability is agreed upon.) The coherence of the system for scientific induction presented in *Theory of Probability* is what struck us the most in *Theory of Probability*, even though we presumably skimmed too fast over this point.

As already noted (with a different twist) in the discussion about Professor Gelman's comments, there is no end to the debate about non-informative priors and, while Professor Zellner's maximal data information prior is an interesting alternative to Jeffreys's, Laplace's and Haldane's solutions, there is no reason to believe the community as a whole will eventually agree upon this point. We obviously appreciate the derivation of this prior based on a specific information criterion developed by Professor Zellner. In a historical perspective, it may well be that the notions of "objective" or "noninformative" are not appropriate for the (Statistics of the) mid-1930s.

The conclusion presented by Professor Zellner reproduces Seymour Geisser's assessment of *Theory of Probability*, for which we are both grateful and in complete agreement.

## 7. CONCLUSION

We are most grateful to the contributors for their lively discussions, which illustrate how influential Jeffreys's ideas still are today. Maybe the most striking aspect in *Theory of Probability* is Harold Jeffreys's intuition that a completely coherent system could be designed for Bayesian analysis, a system upon which we are still building today.

## ACKNOWLEDGMENTS

This reply was written after the first author attended both the O-Bayes 2009 and MaxEnt 2009 conferences. He is grateful to the speakers in the special Harold Jeffreys's *Theory of Probability* session at O-Bayes 2009 and to the participants from those conferences who offered comments on the paper or simply support for the project.